State and Network Structures of Stock Markets around the Global Financial Crisis

Jae Woo Lee·Ashadun Nobi

**Abstract** We consider the effects of the 2008 global financial crisis on the global stock market before, during, and after the crisis. We generate complex networks from a cross-correlation matrix such as the threshold network (TN) and the minimal spanning tree (MST). In the threshold network, we assign a threshold value by using the mean and standard deviation of cross-correlation coefficients. When the threshold is equal to the mean of these coefficients, we observe a giant cluster composed of three economic zones in all three periods. We find that during the crisis, the countries in the Asian zone were weakly connected and those in the American zone were tightly linked to the countries in the European zone. At a large threshold, the three economic zones were fragmented. The European countries connected tightly, but the Asian countries bound weakly. The MST constructed from the distance matrix. In the MST, France remained a hub node in all three periods. The size of the MST shrank slightly during the crisis. We observe a scaling relation between the network distance of nodes from the central hub (France) and the geometrical distance. We observe the topological change of the financial network structure during the global financial crisis. The TN and MST are complementary roles to understand the connecting structure of financial complex networks. The TN reveals to observe the clustering effects and robustness of the cluster during the financial crisis. The MST shows the central hub and connecting node among the economic zones.

**Keywords** Stock market· Minimal spanning tree · Threshold network · Crisis

———————————————

J. W. Lee (✉)

Inha University, Incheon, Republic of Korea

e-mail: jaewlee@inha.ac.kr

A.Nobi

Noakhali Science and Technology University, Sonapur, Bangladesh

e-mail: ashadunnobi_305@yahoo.com

# 1 Introduction

The 2008 global financial crisis (henceforth crisis) was one of the most important phenomena in financial markets. It is very difficult to predict such a crisis. There are several methods to study financial markets(He and Deem 2010; Huang et al. 2009; Kantar et al. 2012; Kumar and Deo 2012; Mantegna 1999; Nobi et al. 2014; Song et al. 2011; Onnella et al. 2003; Vandewalle et al. 2001; Wilinski et al. 2013; Zheng et al. 2013; Zhao et al. 2016; Hui and Chan, 2015). The network approach has emerged as an important technique to describe its static and dynamic properties (Baba and Packer, 2009; Eryighit and Eryigit, 2009; Lin et al. 1994; Namaki et al. 2012; Onnela et al. 2005; Qiu et al. 2010; Sienkiewicz et al. 2013; Brida et al. 2016; Nobi and Lee, 2016; Wang and Xie, 2016). Network analysis based on cross-correlations is limited not only to return time series data on stock prices, but also to quasi-synchronously recorded time series of global indices worldwide(Namaki et al. 2012; Onnela et al. 2005; Nobi et al. 2015). In an interdependent economic world, relationships between financial entities have emerged as an important area of study. Recently, correlation based network analysis of global financial indices has been applied successfully in analyzing the structural change of the global financial network(Eryighit and Eryigit, 2009; Lin et al. 1994; Namaki et al. 2012; Onnela et al. 2005; Qiu, Zheng and Chen, 2010; Sienkiewicz et al. 2013; Wang and Xie, 2015). With the cross-correlation network method, one may observe the properties of eigenvalues, eigenvectors, and the threshold network (TN), generated by assigning a value of threshold (Lin et al. 1994), the minimal spanning tree (MST)(Onnela et al. 2003), and the planar maximally filtered graph (PMFG)(Song et al. 2011).

    Onnela et al. applied correlation network techniques to observe the structural transition of the financial network during the crisis, for a local market (Onnella et al. 2003). Vandewalle et al.(2001) observed the power law of the degree distribution in the MST of the U.S. stock market. Qiuet al.(2010) reported the static and dynamic financial networks based on the average threshold of cross-correlations. Kumar and Deo (2012) applied network analysis to the global financial market based on the absolute threshold of cross-correlation coefficients. They found the European indices were tightly linked during the crisis, at a high threshold. Namaki et al.(2012) observed the threshold network on the Tehran stock market and the Dow Jones Industrial Average. They found scale-free threshold networks on a restricted range of the threshold. Huang et al.(2009) also applied a threshold network analysis on the Chinese stock market. Kantar et al.(2012) observed that in the MST, Turkey's companies listed on the country's stock market were not influenced by the crisis. Eryigit and Eryigit(2009) reported the MST and PMFG in the world stock market. They observed that the French FSBF 250 is a hub node in the MST. Zheng et al.(2013) reported observing the MST and hierarchical network (HN) in the worldwide finance and commodities markets.

    In this article, we consider the network structures of global financial indices generated from cross-correlation coefficients worldwide. We observe that after the Lehman Brothers bankruptcy in September 2008, all global indices experienced high fluctuations. We consider daily data of each country's stock index to calculate the return, observing that the smoothing

volatility for all indices increased sharply until mid-October that year, and then decreased. The trend of volatility after the crisis differs greatly from that during the crisis. The novelty of this article lies in the extraction of information on global indices after the financial crisis and in the comparison of these indices with those before and during the crisis. We propose a method to construct threshold networks from the cross-correlation coefficients of the index change before, during, and after the crisis. We report that regional interaction during the crisis was stronger than in all other periods. Typically, a physical map merely presents the location of a country and its neighbors on the globe but does not explain how two strong countries interact in the world stock market. However, a financial map of assets such as the MST constructed from the distance matrix highlights the interactions among not only regional countries but also among the non-regional ones. In this article, we construct a financial map not only to visualize the relationships between the global indices according to their economic zones but also to examine the economic reasons for interactions between two non-regional clusters. We observe the obvious topological changes of the MST before, during, and after the crisis. Moreover, we study the relationship between geometrical distances and network distances and show a scaling relation.

The paper is organized as follows: in Sect. 2, we discuss the set of investigated financial data and we introduce the volatility trend and the statistical properties of the indices. The threshold networks are discussed in Sect.3, and the minimal spanning tree is described in Sect. 4. In Sect. 5, we present our concluding remarks. In Appendix 1, we give the full name of the global stock market.

## 2 Data and Methodology

We analyze the time series of the daily closing stock prices of global indices located worldwide from June 2, 2006 to December 30, 2010. The total number of the indices is 35 (the full names of the stock markets are provided in Appendix). Of the 35 markets, 17 belong to the European economic zone, and 13 to the Asian-Australian economic zone; only five indices belong to the American economic zone. We obtained the entire data set from the Bloomberg web site[1] and divided the data into three periods on the basis of volatility. We observed that mean volatility during June 2, 2006 to November 30, 2007 was the minimum; we consider this period as "before the crisis." The period from December 3, 2007 to June 30, 2009 is considered as the crisis period due to high mean volatility in all indices (Nobi et al. 2013). This period witnessed the Lehman Brothers bankruptcy that ignited the global financial crisis (Baba & Packer, 2009). Before 2008, the subprime mortgage crisis hit the U.S. housing market, the accumulated pressure of which burst partly in 2007 and erupted finally in 2008. We consider the period July 1, 2009 to November 30, 2010 as "after the crisis" where the mean volatility of some developed markets returned to normal state.

---

[1] http://www.bloomberg.com/

Let the closing stock price of an index be $I_i(t)$ at time *t*. For each window, we calculate the returns of the price fluctuation of each index as

$$R_i(t) = [ln\, I_i(t) - \frac{ln\, I_i(t-1)]}{\sigma_i}, \qquad (1)$$

where $\sigma_i$ is the standard deviation of price changes. Then, we calculate the cross-correlation coefficients as

$$C_{ij} = <r_i(t)r_j(t)> - <r_i(t)><r_j(t)>, \qquad (2)$$

between the return time series of the index *i* and *j*.

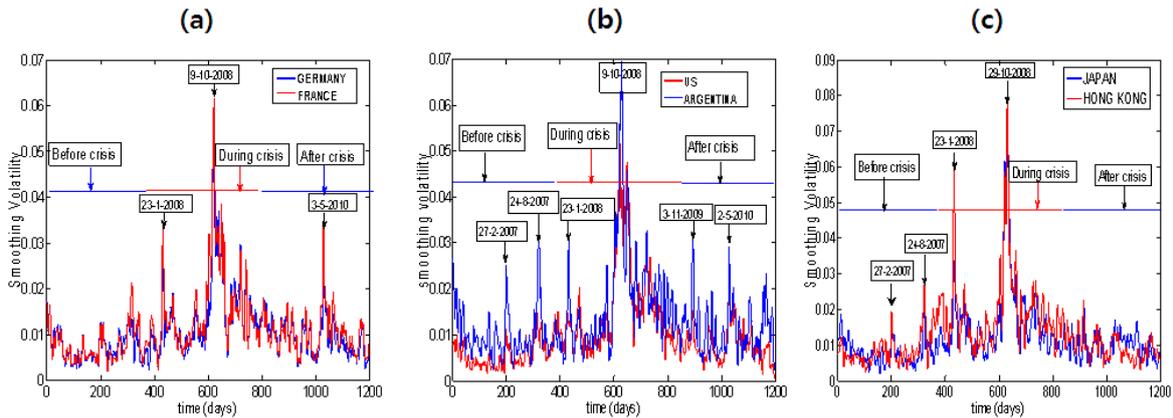

**Fig. 1** Smoothed volatility of main countries during the global financial crisis: (a) Germany and France, (b) the United States and Argentina, and (c) Japan and Hong Kong. The volatility shows synchronized behavior.

The absolute value of the return is known as volatility. To observe a clear trend in volatility, we smooth the volatility using a locally weighted scatter plot smoothing (LOWESS).This data analysis technique produces a "smoothened" set of values from a time series characterized by a scatter plot with a "noisy" relationship between the variables. The LOWESS regression, introduced by Cleveland (1981), is based on a smoothing procedure that pays greater attention to local points (Lin, Engle & Ito 1994). The smoothened value of the volatility *y*, corresponding to a data point *x*, is obtained on the basis of the data points around it within a band of certain width. We group the five nearest neighboring values as a local region to calculate each LOWESS value. In Fig. 1, we observe that the trend of volatility for regional indices is almost similar. Before the crisis, the smoothing volatility shows no sharp transition, indicating a stable market. The sharp change in volatility is clearly visible during the crisis, indicating an unstable market. The highest peak of smoothing volatility for Asian indices is observed at the end of October 2008, while for European and American indices this is observed at around the beginning of October 2008. This indicates a similar trend in volatility for regional indices. After the crisis, the Asian market showed the greatest stability, as shown in Fig. 1(c), while European and American indices still experienced small peak values in volatility. Due to the sharp change in the volatility, the mean volatility of all indices is higher than in other periods during the crisis, as shown in Table 1.

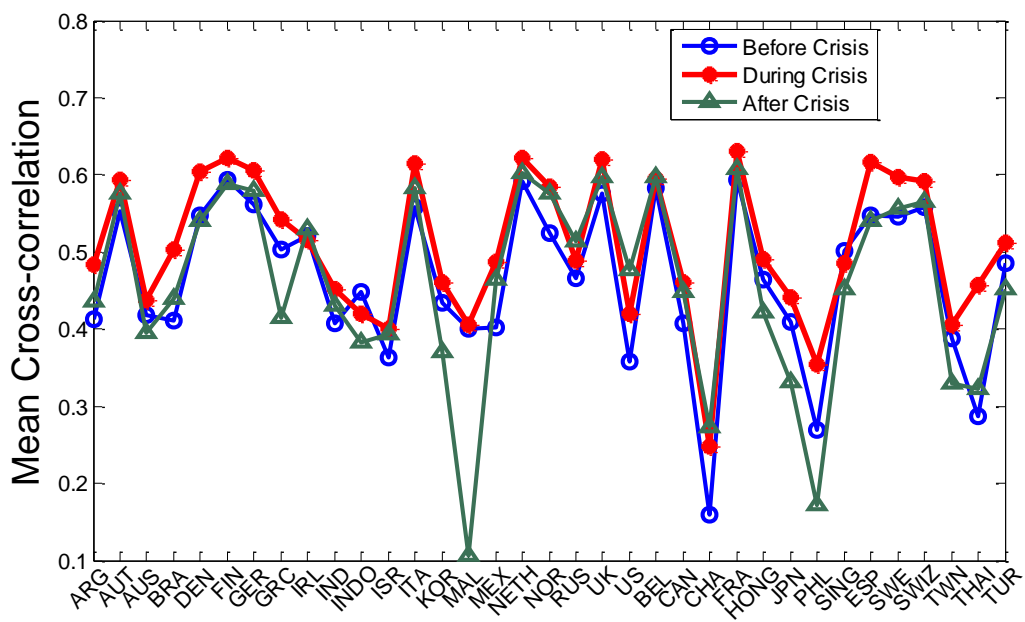

**Fig. 2** The mean cross-correlation coefficients of each index refer to before (green open circle), during (red filled circle), and after (green triangle) the crisis.

In Fig. 2, we present the mean cross-correlation coefficient of each index for the three periods. The mean cross-correlation coefficients values in almost all countries during the crisis are larger than those before and after the crisis. In the European zone, the countries are tightly correlated in all three periods. However, in the Asian and American zones, the values of the cross-correlation are less than those of the European zone. For example, the mean correlations in Malaysia, China, and the Philippines are less than 0.3 in almost all periods. Moreover, the small values of the cross-correlation are maintained in these three countries after the crisis. Before the crisis, we observe two dominant peaks (February and August 2007) of smoothing volatility in the Argentinean market (Fig. 1(b)) and in the Japanese and Hong-Kong markets(Fig. 1(c)). These large fluctuations were induced by the U.S. mortgage crisis. During the period of the crisis, we observe a large peak in January in all markets. On October 9, 2008, before the bankruptcy of Lehman-Brothers, the largest peaks of smoothed volatility are observed in all markets, as shown in Fig. 1. There were big aftershocks on October 28, 2008 in all markets. A big after-peak is observed in the Hong-Kong market in Fig. 1(c). After the crisis, a big peak in the smoothed volatility corresponding to the ESD in the European market in Fig. 1(a) is observed.

In Table 1, we observe that the mean, skewness, and mean volatility of the cross-correlation coefficients during the crisis show high values compared to the other two periods. However, the standard deviation during the crisis is less than that of the other two periods. In Table 1, the mean cross-correlation (0.507) during crisis is higher than before (0.464) and after crisis (0.458). It implies that markets are strongly interacted with each other during crisis. However, before and after crisis, markets are almost similar state in a sense of mean correlation. The high mean cross-correlations during the crisis indicate strong interaction among the markets when the crisis finally hit the world stock market. The lower standard deviation during crisis indicates that the distribution of cross-correlation coefficient is narrower than other periods. It is due to high cross-correlation coefficients almost all indices in this period. Skewness is a measure of the lack of symmetry. The higher positive skewed (0.882) during crisis is due to strong interaction of the European and American indices. We observe heavy tail (5.72) of the correlation distribution after crisis. The distribution of the cross-correlation coefficients becomes narrower in the period during the crisis, compared to the other two periods. However, the distribution during the crisis is more positively skewed than during other periods. The volatility during the crisis shows high values, which indicates high fluctuations in the world market in this period.

**Table 1** Comparison of statistical properties before, during, and after the crisis. During the crisis, the mean cross-correlation and mean volatility increase, but standard deviation and skewness decrease.

| Period | Mean Cross-Correlation | Standard Deviation | Skewness | Kurtosis | Mean volatility |
|---|---|---|---|---|---|
| Before | 0.464 | 0.193 | 0.727 | 4.95 | 0.0088 |
| During | 0.507 | 0.176 | 0.882 | 5.42 | 0.0163 |
| After | 0.458 | 0.206 | 0.722 | 5.71 | 0.0094 |

## 3 Threshold Networks in the World Stock Market

We generate threshold networks assigning a threshold value θ of the cross-correlation coefficients(Huang, Zhuang & Yao, 2009; Nobi et al. 2013). In the threshold network, a node

(V) represents a distinct index of a country and a link (E) represents the connection between two indices weighted by the cross-correlation value of the return time series between two indices. We calculate average cross-correlation coefficients and standard deviations in Table 1. We specify a certain threshold θ, $-1 \leq \theta \leq 1$, from the cross-correlation coefficients. If the correlation coefficient $C_{ij}$ between two countries is greater than or equal to θ, we add an undirected link connecting the nodes *i* and *j*. Therefore, different thresholds define networks that have the same set of nodes but different sets of links. In Fig. 3, we construct the threshold networks with the threshold as the average value of the cross-correlation coefficient, plus the integer multiplication of the standard deviation. Because the distributions of the cross-correlation depend on the observing time period, we require a fair criterion to generate the threshold network. If we assign the threshold as the sum the average and standard deviations of the cross-correlation coefficients, the connection structures of the threshold network are similar, but with different topological properties. If we assign the threshold as the absolute value of the cross-correlation coefficients, we observe very different connecting structures. For example, at a high absolute threshold, the largest threshold network during the crisis becomes very small because the average of the cross-correlation coefficient is larger than in the other periods.

At the threshold $\theta = \overline{C_{ij}}$, one large cluster appears in all periods where the three regional sub-clusters(Asian, European, and American) are obviously visible. During the crisis, as opposed to the other periods, the Asian cluster was loosely connected to the European cluster, as seen in Fig. 3b. Japan and Russia were also connected during the crisis. In addition, during the crisis, the world market was separated into two main components, the European-American economic zone and the Asian economic zone. In the divided economic zone, the indices are more tightly correlated to one another. The Asian market decoupled from the risky American market during the global financial crisis. However, before and after the crisis, the Asian and European economic zones were connected by many links.

We measure the characteristic path length $\bar{l}$ between the nodes in the clusters. The characteristic path length, or the average shortest path length, in a cluster is defined by (Onnela et al. 2003)

$$\bar{l} = \frac{2}{N(N-1)} \sum_{\substack{i,j \\ i<j}} l_{ij}, \qquad (3)$$

where $l_{ij}$ is the shortest path length between the node *i* and *j*. At the threshold, $\theta = \overline{C_{ij}}$, we obtain the characteristic path lengths in the largest cluster as $\bar{l} = 1.65$ before the crisis, $\bar{l} = 2.38$ during the crisis, and $\bar{l} = 1.69$ after the crisis. The large value of the characteristic path length during the crisis is supported by a weak connection between the Asian and the European economic zones. However, these two zones recovered their strong connections after the crisis. The American economic zone, however, was always bound tightly to the European economic zone.

Interesting features are observed at the threshold, $\theta = \overline{C_{ij}} + \sigma$, where regional clusters are distinctly visible. Before the crisis, all nodes of the European zone appear to constitute a large cluster, while the Asian indices form a smaller cluster with eight nodes. However, the American zone connects to a fully connected cluster. The formation of three regional clusters corresponds to the world economic zones. The characteristic path lengths of the each cluster before the crisis are $\bar{l} = 1.2$ for the American cluster, $\bar{l} = 1.40$ for the European cluster, and $\bar{l} = 1.53$ for the Asian cluster (which is weakly connected in comparison to the other clusters). During the crisis, the European indices bridge a fully connected cluster from which Russia and Turkey are disconnected. The Asian indices construct a linear cluster keeping South Korea at the center of the cluster, which may play an important role in rescuing this economic zone from the crisis. We obtain the characteristic path lengths during the crisis as $\bar{l} = 1.30$ for the American cluster, $\bar{l} = 1.19$ for the European cluster, and $\bar{l} = 2.0$ for the Asian cluster. Because the Asian cluster appears like a linear cluster, its characteristic path length appears to increase. After the crisis, the European and American clusters with the characteristic path length $\bar{l} = 1.68$ combine through the United States to form one large cluster. On the other hand, the Asian cluster with mean cluster size $\bar{l} = 1.66$ forms a linear connecting cluster with the nodes Singapore, Hong Kong, Australia, and Japan.

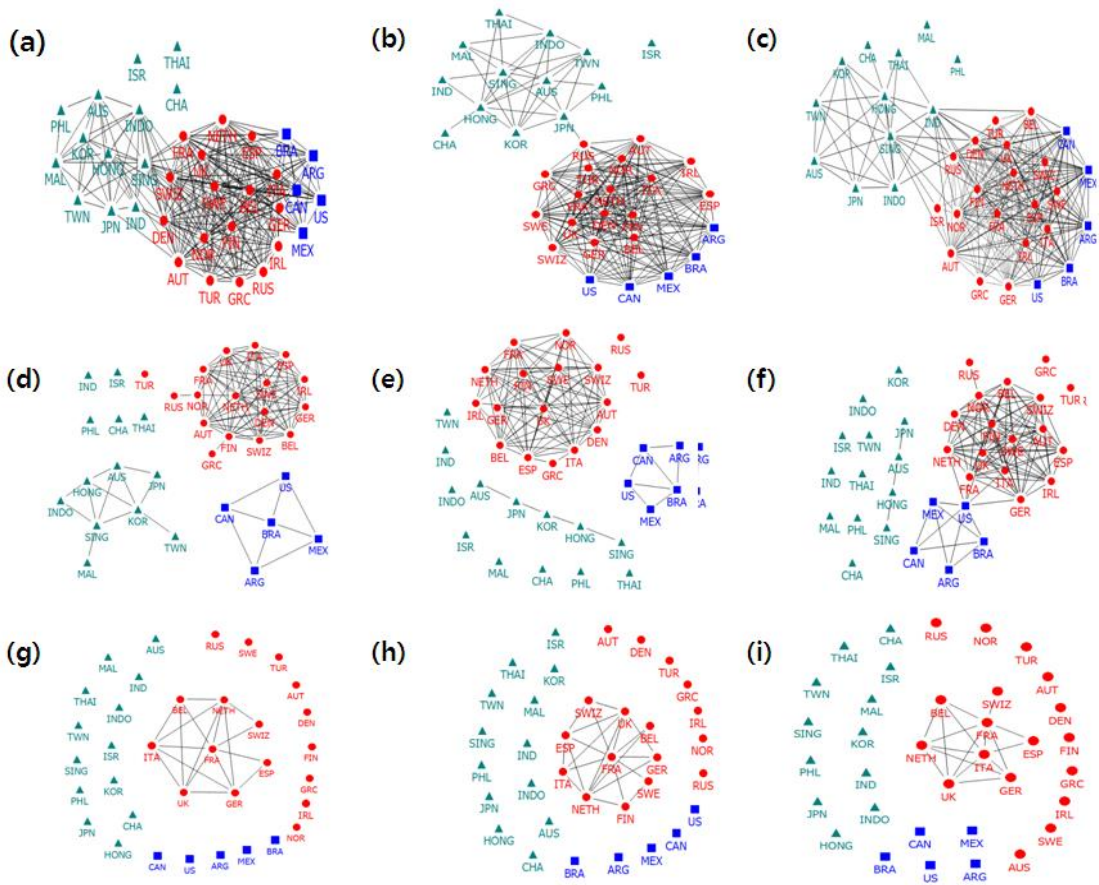

**Fig. 3** The threshold networks are generated from different values of the threshold for before, during, and after the crisis. The abbreviated name below the node indicates the name of the country (see Appendix A). We use different symbols for each economic zone, e.g., the Asian economic zone is a green triangle, the European economic zone is a red circle, and the American economic zone is a blue square. We observe a large cluster in all periods. We present the network clusters: (a) before($\theta$=0.464), (b) during($\theta$=0.507), and (c) after($\theta$=0.458) the crisis at the threshold, $\theta = \overline{C_{ij}}$; (d) before($\theta$=0.657), (e) during($\theta$=0.683), and (f) after($\theta$=0.664) the crisis at the threshold, $\theta = \overline{C_{ij}} + \sigma$; and (g) before($\theta$=0.85), (h) during($\theta$=0.859), and (i) after($\theta$=0.87), the crisis at the threshold, $\theta = \overline{C_{ij}} + 2\sigma$. At a low threshold, we observe a large cluster with a few isolated nodes. At a high threshold, however, several independent clusters with many isolated nodes exist.

At the threshold $\theta = \overline{C_{ij}} + 2\sigma$, only the European economic zone forms a cluster. Eight countries exist in the cluster during all the periods. We obtain the characteristic path lengths of the European cluster as $\bar{l} = 1.35$ before the crisis, $\bar{l} = 1.51$ during the crisis, and $\bar{l} = 1.39$ after the crisis. The core European countries interact very strongly across all periods.

We determine the global efficiency for unconnected vertices in the network. The global efficiency is defined as $E = \frac{1}{N(N-1)/2} \sum_{i>j} \frac{1}{l_{ij}}$, where $l_{ij}$ is the shortest path length from node $i$ to $j$ and $N$ is the total number of nodes in the network (Latora & Marchiori 2001; Yan, Xie & Wang 2014). The global efficiency for the successive thresholds mentioned is 0.607, 0.2174, and 0.038 for before crisis, 0.5739, 0.1847 and 0.056 for during crisis and 0.6384, 0.2431 and 0.0378 for after crisis. Like characteristic path length, the global efficiency during crisis is lower than the other periods. It may be due to the less communication of Asian zone with European and American Zone during crisis.

To characterize how compact or tight the sub-graph of the threshold network is, we use the concept of sub-graph intensity(Onneal et al. 2005), which permits us to characterize the interaction patterns within communities. The sub-graph geometric intensity is expressed as

$$I_g = \left(\prod_{(ij) \in E_g} w_{ij}\right)^{1/E_g}, \qquad (4)$$

Where $E_g$ is the total number of links in the sub-graph and $w_{ij}$ is the weight of the links. At the threshold $\theta = \overline{C_{ij}}$, the intensity of the largest cluster, $I_g = 0.669$, during the crisis is slightly bigger than $I_g = 0.623$ before the crisis and $I_g = 0.639$ after the crisis. This indicates that global indices were tightly connected during the crisis. The intensity at the threshold, $\theta = \overline{C_{ij}} + \sigma$, increases during the crisis. The mean intensities of the European, Asian, and American clusters are 0.77, 0.707, and 0.715, respectively, before the crisis, whereas these values during the crisis are 0.811, 0.713, and 0.745, respectively. After the crisis, the European and American clusters combine to emerge as the largest cluster. The strength, $I_g = 0.796$, is weaker than the intensity of the European economic zone during the crisis. Besides, the intensity of the Asian cluster, $I_g = 0.691$, decreases after the crisis, implying that the Asian indices showed negative tendencies to form communities. The European zone that forms a cluster only at the threshold $\theta = \overline{C_{ij}} + 2\sigma$ is clustered more tightly during the crisis. The intensities of the European cluster at the threshold, $\theta = \overline{C_{ij}} + 2\sigma$, are 0.88, 0.89, and 0.905, before, during, and after the crisis, respectively. We can conclude that during the crisis, European communities interacted strongly, while Asian and American communities interacted loosely.

In recent years, the detection of community and characterization has gained increasing research attention (Fortunate, 2010; Newman, 2006; Newman, 2011). The subsets of vertices, which are densely connected internally while being sparsely connected to the rest of the network, are called communities or modules. Networks such as the karate club of Zachary (Zachary, 1977), collaborations between early jazz musicians of Glesier and Danon(2003), and email contacts between students(Ebel et al. 2002) are characterized by modular structure. To detect modularity, the most successful method is the one proposed by Newman that allows the comparison of different network partitions. Our focus is to detect modularity of the global financial networks using the Newman optimization method to compare the modularity structure before, during, and after the global financial crisis. Here, we consider the community of the global financial indices according to geographical location. For a given network of *N* vertices, the modularity is expressed by

$$Q = \frac{1}{4m}\sum_{ij}(A_{ij} - \frac{k_i k_j}{2m})\delta_{c_i,c_j}, \qquad (5)$$

where $A_{ij}$ are the elements of the adjacency matrix with values 0 and 1, $k_i k_j/2m$ are the expected number of edges between vertices *i* and *j* if the edges are placed at random, $m = \frac{1}{2}\sum_{ij}A_{ij}$ is the total number of links in the network, $\delta_{ij}$ is the Kronecker delta symbol, and $c_i$ is the community containing node *i*. All vertices belong to the same community when Q=0. However, higher values of Q indicate strong community structure. We apply the Newman algorithm to detect modularity for the threshold network generated from the average value of cross-correlation coefficients and also for the MST before, during, and after the crisis. We obtain the modularity of the threshold network at the threshold, $\theta = \overline{C_{ij}}$, as Q=0.213, 0.245, and 0.178 for before, during, and after the crisis, respectively. The modularity was seen to increase during the crisis, but decrease after the crisis. The high modularity implies that during crisis the nodes were strongly connected to one another and consisted of communities such as economic zones.

**4 Minimal Spanning Tree (MST)**

The MST is also constructed by calculating the distance matrix of the indices(Kumar & Deo 2012; Nobi et al. 2013; Onnella et al. 2003). The distance between countries *i* and *j* is defined by $d_{ij} = \sqrt{2(1 - C_{ij})}$, where $d_{ij}$=0 if index *i* and *j* are perfectly correlated, and $d_{ij}$=2 if index *i* and *j* are perfectly anti-correlated. The MST is built following Kruskal's algorithm in order to find the *N-1* most important correlated pairs of the indices among the *N(N-1)/2* possible pairs. We construct the MST in order to visualize a complex financial network before, during, and after the global financial crisis. In Fig. 4, we observe three regional clusters, i.e., European, Asian, and American, in all periods, with France always placed as the central node in the MST. France has become an international business center over the last decade, emerging as the fourth "legend" in the world of foreign investment. More than 18,000 subsidiaries of foreign companies are located there, connecting Paris to the rest of the world. Paris holds the second position globally in terms of hosting the headquarters of these companies, after Tokyo, and ahead of London and New York. As a result, it is a hub for many national corporations and consequently occupies the center hub of the MST.

Before the crisis, Singapore was the hub of the Asian cluster, as seen in Fig 4(a). Singapore is one of the most open and competitive markets worldwide. The 2011 World Bank Ease of Doing Business Index ranks Singapore as the best country in the world for business[2]. Singapore has earned its reputation as Asia's financial center by virtue of three important pillars. First, Singapore is well known as the home of the Asian domestic market (ADM); by borrowing loans from outside Asia as well as from neighboring countries, it meets the increasing demand for loans for the rapidly expanding Asian countries. Second, it is the fourth-largest foreign exchange center in the world. Third, the Singapore International Monetary Exchange (SIMEX) has grown to become one of the major futures and options exchanges in the world[3]. SIMEX has set up dynamic derivative markets that thrive principally on instruments traded in the ADM and the foreign exchange market. As a result, it sits at the center of the Asian cluster in the MST and is directly connected to the European cluster. China, one of the largest and most populated countries in the world, does not carry any significant role in the MST and in the threshold network.

The United States emerges as a hub of the American cluster and it was directly connected to the European cluster before the crisis. The U.S. economy is the engine of the global economy, with the U.S. markets driving all other markets globally. An important role in financial markets is explained by foreign asset prices. On average, about 26 percent of the movements in European financial assets are attributable to developments in U.S. financial markets, while about 8 percent of U.S. financial market shifts are caused by European developments[4]. The significance of U.S. markets is found particularly in equity markets; for instance, movements in U.S. stock prices trigger corresponding changes in the Euro area, with more than 50 percent of U.S. market developments being reflected in Euro area stock prices. By contrast, European equities have an insignificant impact on their American counterparts. This confirms the central role that the U.S. equity markets play in world stock markets. As a result, the United States interlinks the European and American cluster in the MST. However, a question arises as to the reason the United States did not emerge as a central hub of the MST. Several factors have decreased the dependence of emerging and other developed markets on the U.S. economy. The world's stock markets have progressed rapidly because of the globalization, and such behavior exhibits more divergence than convergence. Second, the Euro has become a much more stable currency than the U.S. dollar, allowing central banks all over the world to reduce their country's exposure to the latter. Third, emerging economies such as China, Russia, and India are taking over the traditionally European and American roles of lending to other emerging nations. These emerging economies are also investing billions of dollars into other nations that have traditionally depended upon U.S. aid for survival.

---

[2] http://www.doingbusiness.org/

[3] Financial Deregulation and Integration in East Asia, NBER-EASE, NgiamKeeJin, Volume 5.

[4] Stephane Dees and Aurther Saint-Guilhem, European Central Bank, Working Paper Series 1034 (1994).

Hong Kong, Singapore, and Japan are regarded as three giants in the Asian financial market. Singapore is popularly known as the "funding center" and Hong Kong as the "lending center." Hong Kong has grown to become one of the most important international financial centers in the world. A rapid and successful movement towards capitalism in Hong Kong was initiated from the 1950s. During the crisis, Hong Kong became a hub of the Asian cluster, as seen in Fig. 4b. The local hub in the MST reflects the fact that Hong Kong played an important role in this economic zone in resisting crisis shocks. Moreover, after the crisis too, Hong Kong remained the hub of the Asian cluster, suggesting that Hong Kong will remain the financial center of the Asian zone in the near future.

In the MST, the Asian cluster is connected to the European cluster through Singapore before the crisis, and Japan during the crisis. In a threshold network, we also observed that during the crisis, Japan was one of the mediators connecting Asian and American clusters at threshold $\theta = \overline{C_{IJ}}$ (see Fig. 4d). Since Singapore and Japan are the legends of the global economy, they interact with the European zone as well as with the rest of the world. However, after the crisis, India has also emerged as an interlink between the European and Asian clusters, as shown in Fig. 4c. Over the last few years, the Indian economy has transformed into a vibrant and rapidly growing consumer market. India becomes a destination for global business and investment opportunities by an abundant and diversified natural resource base, sound economic, industrial, and market fundamentals, and highly skilled and talented human resources. Armed with a population of more than a billion people, India is now the 11th largest economy in the world[5]. In the near future, it can be seen as another legend in the global economy. Hence, in the MST, the interaction of the Asian cluster with the European cluster through India appears reasonable. On the other hand, the United States is displaced from the hub of the American cluster and its direct interaction with the European cluster during the crisis is disconnected (see Fig. 4b). However, after crisis, it again strongly interacts with the European cluster just as before the crisis (see Fig. 4c). The network structure in the MST after the crisis (see the position of Japan, Korea, Singapore, Hong Kong, and the United States) is almost similar to before the crisis.

---

[5] http://www.worldbank.org/

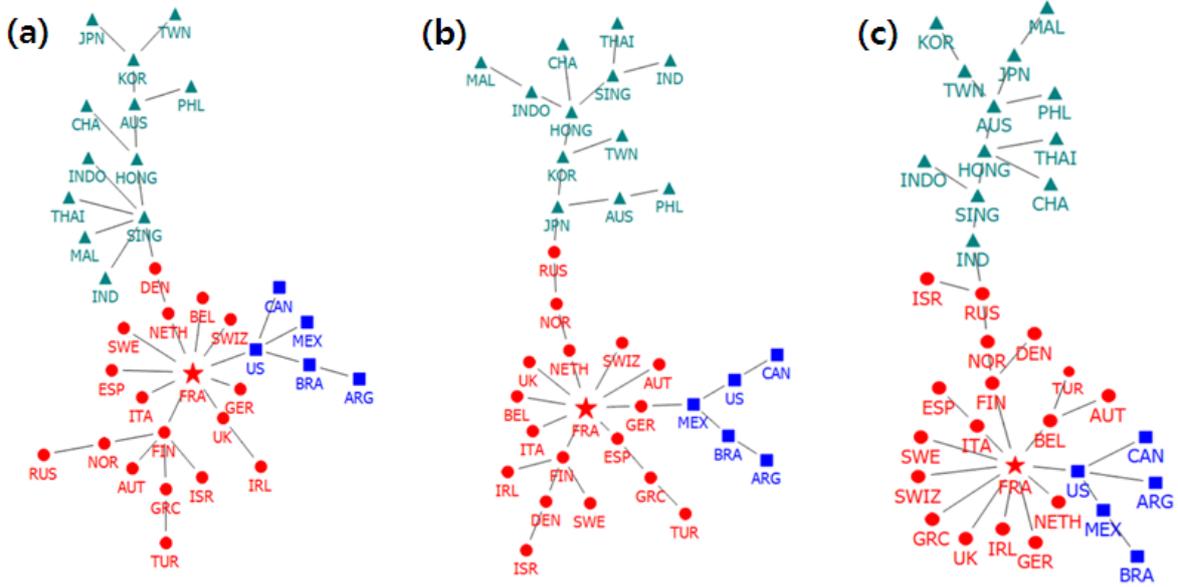

**Fig. 4** The MSTs obtained at three periods: (a) before the crisis, (b) during the crisis, (c) after the crisis. Three regional clusters are very visible. France remains the central hub.

We measured the modularity by Eq. (3) in the MST, obtaining Q=0.313, 0.494, and 0.224, for before, during, and after the crisis, respectively. The modularity value during the crisis is the highest, but after the crisis, it becomes the lowest. During the crisis, the MST forms more modules than in the other periods. The size of the tree characterizes the average tree length in the MST. We define the average tree length as(Onnela et al. 2003)

$$L(t) = \frac{1}{N-1}\sum_{<i,j>} d_{ij}^{MST}, \qquad (6)$$

where $N$ is the total number of nodes in the tree and $d_{ij}^{MST}$ is the distance between two neighboring nodes $i$ and $j$. We calculate the average tree length $L$ in the MST. The measured average tree lengths are $L = 0.73, 0.70$, and $0.72$, for before, during, and after the crisis, respectively. During the crisis, the size of the MST is smaller than during the other periods, which indicates that the indices behaved in a tightly correlated manner. This kind of strong interaction holds true for the regional cluster (though not for regional clusters) that we observed from the threshold network and MST network.

In Fig. 5, we consider the relation between the real geometric distance and the network distance from the central hub (France). The geometric distance measures distances from Paris to the cities of the target countries. The network distance, however, corresponds to the shortest path length from the hub node (France) on the MST. We observe the power law behavior $d \sim L^\alpha$ of the network distance on the geometric distance. The solid lines in Fig. 5 represent the best fits to a scaling relation. The scaling exponent α = 0.69(12) during the crisis is bigger than the exponent α = 0.57(11) before the crisis. This result is in complete agreement with the observation that the network distance from the central hub to Asian and American cluster increased during the crisis. However, after the crisis, we obtain the exponent α = 0.72(14), which is almost similar to that during the crisis, if we consider the error bar. This is also in full agreement with the observation that some countries continue to pass through the crisis period. Moreover, the network distance from the central hub to some nodes (Korea, Japan, Australia, and the Philippines) in the Asian cluster, as well as to the European cluster, increases abruptly after the crisis, which creates a larger slope and consequently, induces a larger exponent. These empirical results suggest that the closer the cities, the stronger the interaction with hub-like nodes. We use Kolmogorov-Smirnov test to compare the distributions of the values of the two data sets in every period. The test rejects the null hypothesis at the 5% significance level. However, the trends of interdependencies between geometrical distance and network distance seem like power law. Since, this interdependency depends on various factors in the global economy, we need to pay more attention and engage in further analysis to discover the economic factors behind these two interdependencies.

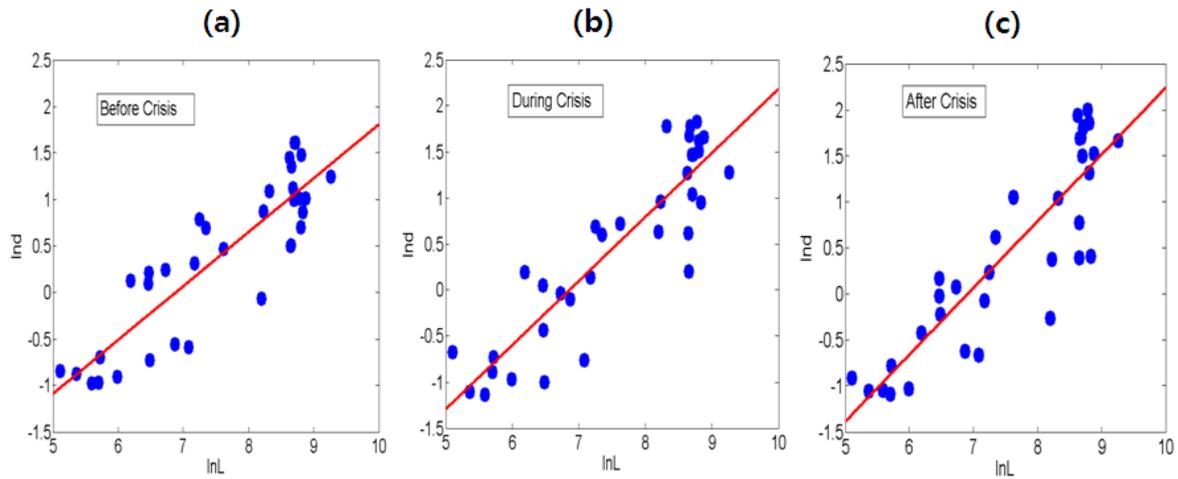

**Fig. 5** Log-log plots of the network distance against the geometrical distance (a) before the crisis, (b) during the crisis, and (c) after the crisis, respectively. Lines show the best fit to a scaling relation $d \sim L^\alpha$. The exponents and goodness of fit are α=0.57($R^2$=0.76), α=0.69($R^2$=0.80), and α=0.72($R^2$=0.76), before, during, and after the crisis, respectively.

**5 CONCLUSIONS**

The financial network we constructed from cross-correlation coefficients of stock index time series yielded information about the reorganization of global indices in the network owing to the2008 crisis. From the threshold network, we observed that the regional interaction among indices of the Asian and American clusters decreased during and after the crisis, while this increased for the European cluster. During the crisis, the countries in the Asian zone were weakly connected to the countries in the European zone. However, the countries in the American zone were tightly bound to countries in the European zone. At a large threshold $\theta = \overline{C_{ij}} + \sigma$, the three economic zones were fragmented. The European countries were tightly connected, but the Asian countries were bound weakly. At a much larger threshold $\theta = \overline{C_{ij}} + 2\sigma$, only the European cluster remained. The MST was also constructed from the distance matrix. We observed the topological change of the network structure during the crisis. In the MST, France was always a hub node in all three periods. We also found out the scaling law from the MST, by observing the relation between the network distance and the geometric distance, and showed that the scaling exponent during the crisis was almost similar to that after the crisis. Further, the configuration of indices on the MST after the crisis was almost similar to that before the crisis, indicating a recovery signal of global indices. The TN and MST are complementary roles to understand the connecting structure of financial complex networks. The TN reveals the clustering effects and robustness of the cluster during the financial crisis. The MST shows the central hub and connecting node among the economic zones.

**Acknowledgements** This research was supported by the Basic Science Research Program through the National Research Foundation of Korea(NRF) funded by the Ministry of Science, ICT and Future Planning(NRF-2014R1A2A1A11051982).

## 6 Appendix

Global Stock Indices are used in the work.

We consider 35 world stock indices. The European economic zone includes 17 countries: France(FRA), Germany(GER), Italy(ITA), the United Kingdom(UK), Espana(ESP), Switzerland(SWIZ), Netherland(NETH), Belgium(BEL), Norway(NOR), Ireland (IRL), Greece(GRC), Finland(FIN), Denmark(DEN), Austria(AUT), Turkey(TUR), Sweden(SWE), and Russia(RUS). In the Asian and Australian economic zone we include 13 countries: Japan(JPN), South Korea(KOR), Singapore(SING), Hong Kong(HONG), Indonesia(INDO), Taiwan(TWN), Malaysia(MAL), China(CHA), Thailand(THAI), India(IND), the Philippines (PHL), Israel(ISR), and Australia(AUS). The number of countries in American economic zone is five, including the United States (US), Canada(CAN), Mexico(MEX), Argentina(ARG), and Brazil(BRA).